\documentclass{ws-ijmpe}

\begin{document}

\markboth{Ali Hanks}{Measuring Bremsstrahlung Photons 
in $\sqrt{s} = 200 \mathrm{GeV}$ p-p Collisions}

\catchline{}{}{}{}{}

\title{Measuring Bremsstrahlung Photons 
in $\sqrt{s} = 200 \mathrm{GeV}$ p-p Collisions}

\author{Ali Hanks, {\it for the
PHENIX\footnote{For the full list of PHENIX authors and
acknowledgements, see Appendix 'Collaborations' of this volume}
Collaboration}}

\address{Columbia University, Nevis Labs,\\
P.~O.~Box 137 Irvington, NY, 10533, United States\\
ahanks@nevis.columbia.edu}

\maketitle

\begin{history}
\received{(received date)}
\revised{(revised date)}
\end{history}

\begin{abstract}
Direct photon production is an important observable in heavy ion collisions, as 
photons are penetrating and therefore largely insensitive to final state effects. 
Measurements of the fragmentation component of direct photon yields in p+p 
and Au+Au collisions will provide important tests of pQCD predictions and 
of predictions for modifications of this component in heavy ion collisions. 
By selecting photons associated with jets on the same side using 
hadron-photon correlations, fragmentation photons can be measured directly. 
\end{abstract}

\section{Introduction}

The phenomenon of jet quenching is one of the key signatures for the presence 
of a hot dense medium in heavy ion collisions. The energy loss of jets as they 
propagate through the medium is thought to be largely due to medium induced 
gluon bremsstrahlung\cite{white}. However, because the radiated gluons 
interact with the medium, the radiation spectrum cannot be measured directly. There 
are several theoretical models that attempt to describe the mechanisms for jet 
energy loss, but all rely on a variety of assumptions including the thickness of 
the medium and the energy of the parton. It would be useful to have a direct 
probe of the radiating parton at all stages in its evolution. In addition to 
gluons, photons are produced through jet fragmentation, and therefore could 
provide just such a probe.

\begin{figure}[ht]
\begin{center}
\psfig{file=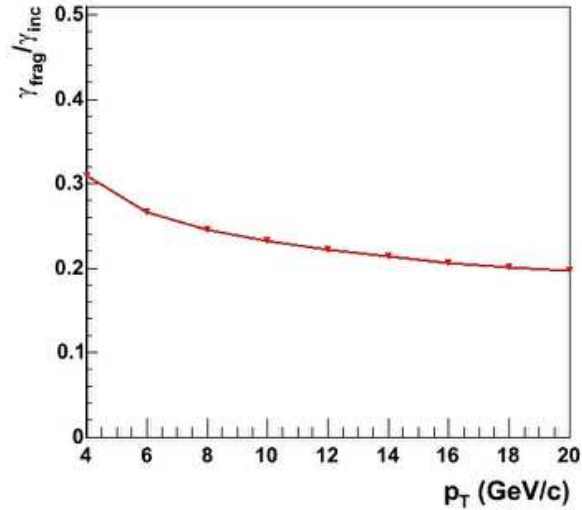, width=3in}
\vspace*{8pt} \caption[]{The ratio of fragmentation photon to inclusive
direct photons using the INCNLO(v1.4) pQCD calculation\cite{incnlo} }
\label{fig:ratio}
\end{center}
\end{figure}

Direct photons provide a powerful probe of final state effects, because once produced 
they do not interact strongly with the medium. In p+p collisions, NLO perturbative 
QDC calculations for the single particle spectra agree well with the data, and the 
ratio of spectra for Au+Au collisions to that for p+p, scaled by the number 
of binary collisions, has already been used as a baseline for understanding final 
state effects on the suppression of high-$p_{T}$ particles\cite{Raa}. At NLO, these 
pQCD calculations  include a significant contribution from photons produced through 
parton fragmentation\cite{incnlo}. At low $p_{T}$ $(p_{T} < 10.0)$, the contribution 
is as much as 20-30\% (Fig.~\ref{fig:ratio}).

In heavy ion collisions the fragmentation component to the inclusive direct photon spectrum 
can be modified. There are two main ways this contribution could be modified. If the 
partons begin radiating bremsstrahlung gluons prior to fragmentation, then the fragmentation 
photon spectrum would be suppressed by as much as 20-40\% for $p_{T} > 3.0 GeV/c$\cite{sup}. 
On the other hand, it is possible that additional photon bremsstrahlung would be induced 
through the interaction of the energetic parton and the medium, leading to an overall 
enhancement of the fragmentation photon spectrum for $p_{T} < 10.0 GeV/c$\cite{enhance}.
This stimulated bremsstrahlung provides direct observation of the scattering of jets in 
the medium, and therefore a way to directly measure the radiation spectrum.

\section{Method}

It is possible to study modifications to the fragmentation component of the direct photon 
spectrum simply by looking at the $R_{AA}$ and studying its deviation from unity\cite{TadaAki}. 
However, because any modifications are likely to be a combination of stimulated bremsstrahlung 
radiation and suppression due to energy loss, and the fragmentation component is 
only a fraction of the total direct photon signal, it is difficult to extract much 
information this way. If instead, the fragmentation component could be measured directly, 
both in p+p and Au+Au collisions, it may be possible to study modifications more closely. 
Because fragmentation photons will be strongly correlated with the high-$p_{T}$ parton that 
produced them, one way of selecting directly for fragmentation photons is to look at hadron-
photon correlations and look at the yield for associated photons on the near side. This 
reduces the background from other direct photons, which should have no correlation with 
the hadrons.

The first step is to obtain the inclusive hadron-gamma per trigger yield. This will include 
the contribution from fragmentation photons, but it will be dominated by decay photons.

\begin{equation}\label{eq:inc}
\frac{1}{N^{h}_{trig}}\frac{dN^{h-\gamma_{inc}}}{d\Delta\phi} = 
\frac{1}{N^{h}_{trig}}\frac{dN^{h-\gamma_{frag}}}{d\Delta\phi}+
\frac{1}{N^{h}_{trig}}\frac{dN^{h-\gamma_{dec}}}{d\Delta\phi}
\end{equation}

\begin{figure}[t]
\centerline{\psfig{file=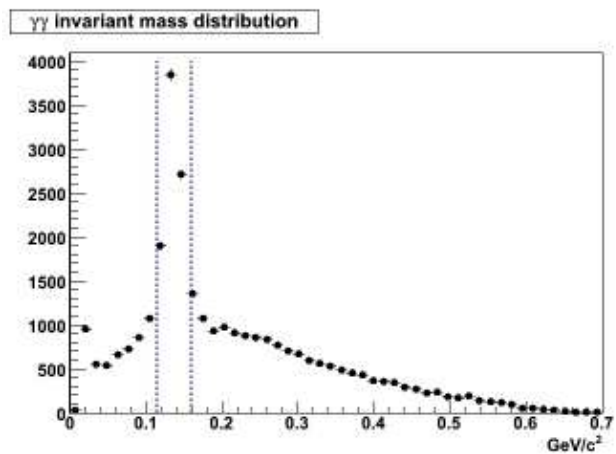,width=3.5in}}
\vspace*{8pt} \caption{Invariant mass distribution for photon pairs. The dashed 
blue lines indicate the mass cut for tagging photons coming from $\pi^{0}$ decays }
\label{fig:mass}
\end{figure}

The dominant source of hadronic decay photons is from $\pi^{0}$ decays. So the next step 
is calculate the invariant mass spectrum for photon pairs and tag those that fall 
within the  $\pi^{0}$ mass peak (Fig~\ref{fig:mass}). There is some inefficiency inherent 
in any tagging method, due to the finite acceptance of any detector, which must be accounted 
for to obtain the true $\pi^{0}$ decay yield. The tagging efficiency is defined as:

\begin{equation}\label{eq:eff}
\varepsilon(\Delta\phi) = \frac{\frac{1}{N^{h}_{trig}}
\frac{dN^{h-\gamma_{tag}}}{d\Delta\phi}}{\frac{1}{N^{h}_{trig}}
\frac{dN^{h-\gamma_{\pi^{0}}}}{d\Delta\phi}}
\end{equation}

To determine this efficiency, a fast Monte Carlo simulation is used to generate $\pi^{0}$s, 
let them decay, and reconstruct them from the resulting photon pairs and study acceptance 
affects. The efficiency will be a function of both $p_{T}$ and $\Delta\phi$. As the 
$\Delta\phi$ between the $\pi^{0}$ and the trigger hadron increases, the possibility 
that one of the decay photons falls outside the acceptance may increase. This will improve 
at higher $p_{T}$, when the initial $\pi^{0}$ is closer to the hadron in $\Delta\phi$. 
To correctly weight for both these effects, the input $\pi^{0}$s are generated with a 
$\Delta\phi$ distribution around a dummy ``trigger'' hadron which is given a random 
$\phi$ and $\eta$ within the detector acceptance. The $\Delta\phi$ distribution is 
determined from the data, as is the input $p_{T}$ spectrum.

\begin{figure}[ht]
\begin{center}
$\begin{array}{cc}
\epsfxsize=2.5in
\epsffile{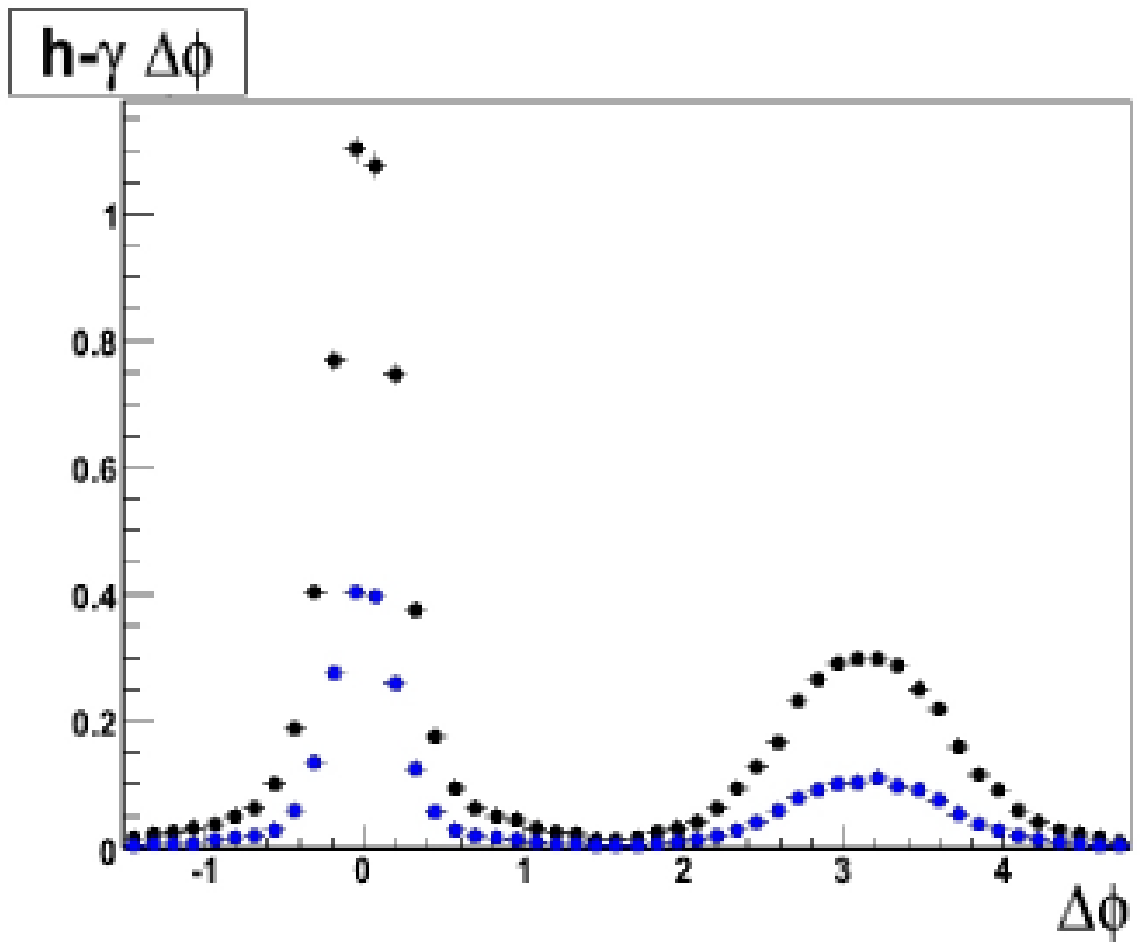} &
\epsfxsize=2.7in
\epsffile{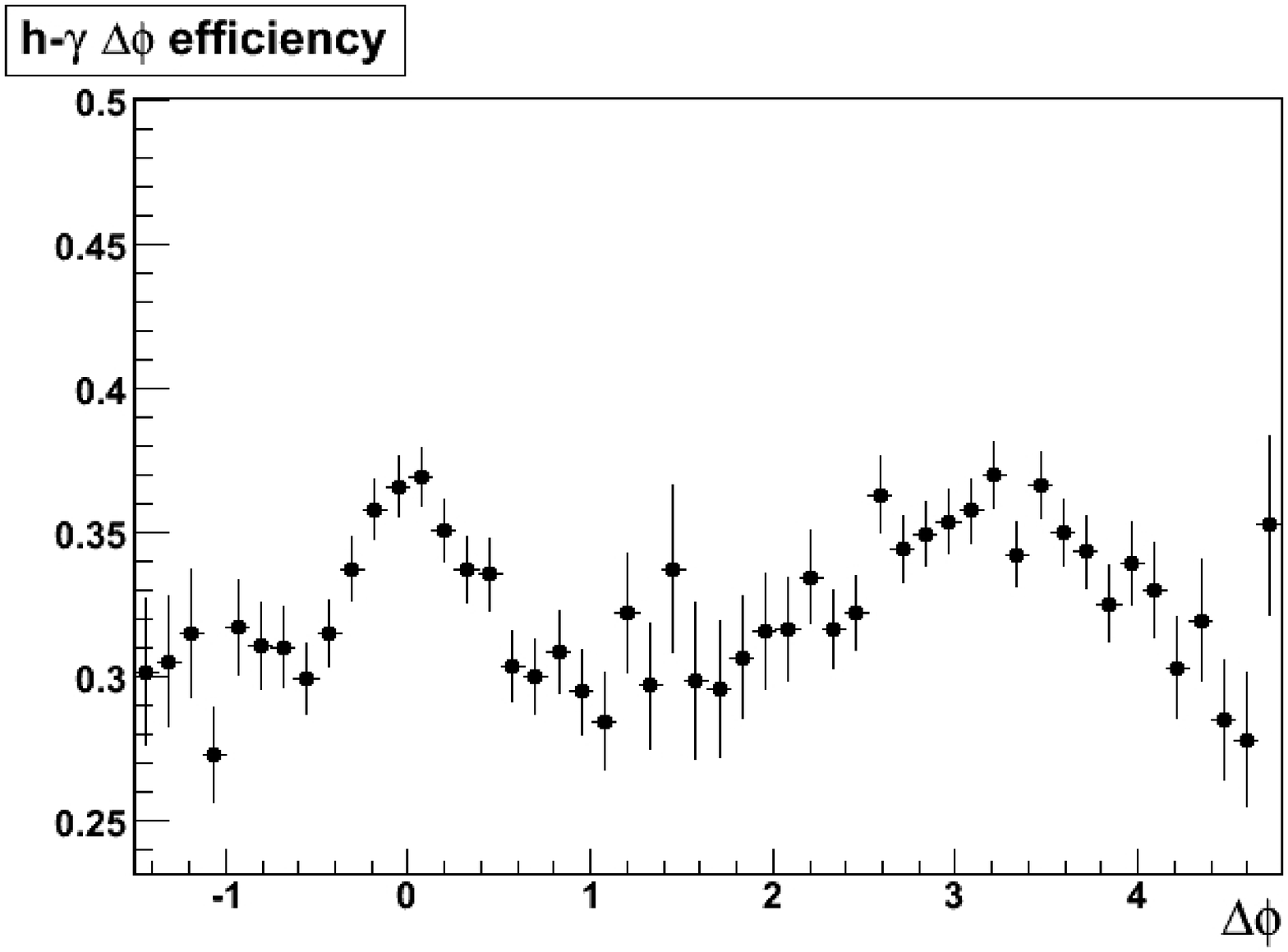} \\
\mbox{\bf (a)} & \mbox{\bf (b)}
\end{array}$
\vspace*{8pt} \caption{{\bf (a)} Simulated hadron-$\gamma$ $\Delta\phi$ distributions 
for photons coming from $\pi^{0}$ decays, requiring one (black) or both 
(blue) of the decay photons to pass the acceptance cuts. 
{\bf (b)} Efficiency for reconstructing $\pi^{0}$ associated with a high 
$p_{T}$ hadron as a function of the $h-\pi^{0}$ $\Delta\phi$. }
\label{fig:fastmc}
\end{center}
\end{figure}

Once the generated $\pi^{0}$ has decayed, the resulting $\Delta\phi$ distribution between 
the trigger hadron and the decay photons can be calculated and broken up into bins in  
$p_{T}$ of the associated photon. The resulting distribution for all photons that 
pass the acceptance cuts is then compared to distribution when both decay photons are 
required to pass the cuts. From these, the tagging efficiency can be calculated. In other 
words, the tagging efficiency is obtained by comparing the case when any single photon 
that has come from a $\pi^{0}$ decay is within the acceptance, and the case when both 
photons from the decay will be detected and correctly reconstructed. Figure~\ref{fig:fastmc} 
shows an illustration of this for $\gamma$ $p_{T}$ from $2.0 - 2.5 GeV/c$.

Once the efficiency is calculated, it can be folded back into the $\pi^{0}$-tagged 
$h-\gamma$ yield to obtain the full $h-\gamma_{\pi^{0}}$ yield. However, while $\pi^{0}$ 
decays are the main source of background, other hadronic decays, for example  
$\eta\rightarrow\gamma\gamma$, are not negligible and must be included in the 
calculation of the full $h-\gamma_{dec}$ yield. Assuming that these additional decay 
contributions have similar $\Delta\phi$ distributions to the $\pi^{0}$, the full decay 
yield is just the $\pi^{0}$ decay yield modulo the ratio of total $h\rightarrow\gamma$ 
decays to  $\pi^{0}\rightarrow\gamma$ decays, as shown in Eq. (3). Folding the tagging 
efficiency in, the total decay yield can now be found in terms of the measurable tagged 
yield.

\begin{equation}\label{eq:dec}
\frac{1}{N^{h}_{trig}}\frac{dN^{h-\gamma_{dec}}}{d\Delta\phi} = 
\underbrace{\frac{N^{}_{h\rightarrow\gamma\gamma}}
{N^{}_{\pi^{0}\rightarrow\gamma\gamma}}}_{R_{h/\pi^{0}}}
\frac{1}{N^{h}_{trig}}\frac{dN^{h-\gamma_{\pi^{0}}}}{d\Delta\phi} = 
R_{h/\pi^{0}}\frac{1}{\varepsilon(\Delta\phi)}
\frac{1}{N^{h}_{trig}}\frac{dN^{h-\gamma_{tag}}}{d\Delta\phi}
\end{equation}

Plugging (\ref{eq:dec}) back into (\ref{eq:inc}) and solving for the fragmentation component 
gives an expression for the final $h-\gamma$ yield for fragmentation photons.

\begin{equation}\label{sub}
\frac{1}{N^{h}_{trig}}\frac{dN^{h-\gamma_{frag}}}{d\Delta\phi} = 
\frac{1}{N^{h}_{trig}}\frac{dN^{h-\gamma_{inc}}}{d\Delta\phi}-
\frac{R_{h/\pi^{0}}}{\varepsilon(\Delta\phi)}\frac{1}{N^{h}_{trig}}
\frac{dN^{h-\gamma_{tag}}}{d\Delta\phi}
\end{equation}

\section{Discussion}

Perturbative QCD calculations match current data very well, and predict that fragmentation 
photons comprise a significant percentage of the total direct photon spectrum at low $p_{T}$. 
This fraction should be large enough to be directly measurable in experiment, once the 
background from decays has been accounted for. Many of the assumptions mentioned in this 
method still need to be studied, such as whether the $\Delta\phi$ distribution for non-
$\pi^{0}$ hadronic decays is really the same shape as for the $\pi^{0}$. For example, the 
$\eta$ has a much larger opening angle, so it is possible that the shape of the yield will 
be modified. These differences may be very small, since $\eta$ decays make up only a small 
fraction of the total decay contribution. However, because the fragmentation photon signal 
is already very small, even such small effects will need to be studied in detail.


\end{document}